\documentclass[twocolumn,prl,aps,superscriptaddress,showpacs]{revtex4}
\usepackage{graphicx}
\begin{document}

\title{Tuning into the Kitaev spin liquid  phase from a new perspective}

\author{Yue Yu}
\affiliation{Department of Physics and Center for Field Theory and Particle Physics, Fudan University, Shanghai 200433, China}
\affiliation{State Key Laboratory of Theoretical Physics, Institute of Theoretical Physics, Chinese Academy of
Sciences, P.O. Box 2735, Beijing 100190, China}
\author{ Long Liang}
\affiliation{Institute of Theoretical Physics, Chinese Academy of
Sciences, P.O. Box 2735, Beijing 100190, China}
\author{Qian Niu}
\affiliation{Department of Physics, The University of Texas at Austin, Austin, Texas 78712, USA}
\affiliation{International Center for Quantum Materials, Peking University, Beijing 100871, China}
\author{ Shaojing Qin}
\affiliation{State Key Laboratory of Theoretical Physics, Institute of Theoretical Physics, Chinese Academy of
Sciences, P.O. Box 2735, Beijing 100190, China}
\date{\today}
\begin{abstract}
We study a spin model on honeycomb lattice with two types of Heisenberg exchange couplings, $J$ and $\tilde J$, where $J$ is for the conventional spin and $\tilde J$ for rotated spin.  When $\tilde J=0$, this is the conventional Heisenberg model. When $J=0$, the system is either in a stripy antiferromagnetic order($\tilde J<0$, ferromagnetic for rotated spin) or  a zig-zag antiferromagnetic order ($\tilde J>0$, antiferromagnetic for rotated spin).  The competition between two ferromagnetic orders  or two antiferromagnetic orders  induces Kitaev's spin liquid phase characterized by the exactly solvable Kitaev model ($J=\tilde J$). Our model can be applied to layered Mott insulators A$_2$IrO$_3$ (A=Li, Na).  For a monolayer of Li$_2$IrO$_3$, we show that it is possible to tune the controlling parameter into the Kitaev spin liquid regime by a link-dependent Rashba spin-orbital coupling. 
\end{abstract}

\maketitle

Quantum spin liquid (SL) is a long-expected new state of matter. Kitaev model on honeycomb lattice gives an exactly solvable theoretical example for the SL. This model  has attracted a lot of interests
because of its exact solvability, Majorana fermion excitation,
non-trivial topological orders  and  amazing abelian and non-ablelian anyons
\cite{kitaev}.  The non-abelian anyon is the key object to design
a device of topological quantum computer \cite{kitaev1}.

 A possible way to realize Kitaev model on optical lattice has been proposed \cite{duan}. Recently, studies for the Kitaev-Heisenberg (KH) model  showed that this peculiar quantum SL is possible to be realized in iridates A$_2$IrO$_3$ (A=Na, Li) because of a strong intrinsic spin-orbital (SO) coupling of the 5$d$ electron of iridium ions\cite{khm,hkm1} .  A phase transition from a stripy antiferromagnetic(AFM) order to the Kitaev SL phase was predicted. Experiments for the iridates, however, found  a zig-zag  AFM order \cite{liu,singh,choi,cao}.   These progresses drive a number of further researches \cite{KY,emf2,RTT,emf1,YD,BLK,dhk1,dhk2,12095100}.

Either the stripy or zig-zag order implies that  the intrinsic SO coupling is too strong for the system to be in the Kitaev SL phase. In order to reach the Kitaev SL phase, it was suggested to reduce the SO coupling with a $c$-axial  pressure for the iridates\cite{singh,sriro}.

\begin{figure}[htb]
\begin{center}
\includegraphics[width=5cm]{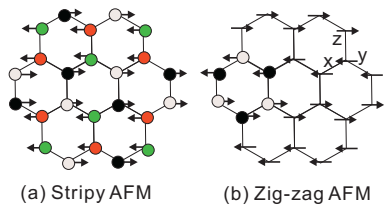}
\caption{ (Color online) (a) The stripy AFM order. Four different colored circles label four sublattices on each of which a rotated spin is defined (See the text)   (b)  The zig-zag AFM order.  }\label{fig1}
\end{center}
\vspace{-0.3cm}
\end{figure}

 In this Letter, we trace out a new perspective  to tune the SO coupling.
We introduce a model on honeycomb lattice with two types of Heisenberg couplings one of which comes from the conventional exchange $J$ and another  from the exchange ($\tilde J$)  for  rotated spin (See Fig. \ref{fig1}(a)). We call this model  the $J$-$\tilde J$ model, which can be mapped to the HK mode with arbitrary couplings by a reparameterization. However, the physical origin of the various phases of the model can be seen more clearly in the $J$-$\tilde J$ model \cite{yuqin}: The stripy antiferromagnetic (AFM) phase in fact is the ferromagnetic(FM) phase of the rotated spin ( $\tilde J<0$ and $J=0$) while the zigzag AFM phase is the AFM of the rotated spin ($\tilde J>0$ and $J=0$). Moreover, when $J=\tilde J$,  the $J$-$\tilde J$ model reduces to the exactly solvable Kitaev model, which reveals an alternative origin of the Kitaev SL:  The competition between the conventional AFM (FM) Heisenberg exchange $J$ and the rotated spin AFM (FM) Heisenberg exchange $\tilde J$ gives birth of the Kitaev SL phase.  Microscopically, the rotated spin Heisenberg exchange comes from a spin-dependent hopping or a link-dependent Rashba spin-orbital (LDR-SO) coupling in  the strongly coupled Hubbard model. For the iridates, we consider the latter.  The correction from a relevant weak LDR-SO coupling may tune Li$_2$IrO$_3$ into the FM Kitaev SL phase.

\noindent{\it  The $J$-$\tilde J$ model and phase diagram. }  
The $J$-$\tilde J$ model is given by
\begin{eqnarray}
H=J\sum_{\langle ij\rangle}{\bf S}_i\cdot{\bf S}_j+\tilde J\sum_{\langle ij\rangle}\tilde{\bf S}_i\cdot\tilde{\bf S}_j\label{JJ}
\end{eqnarray}
where ${\bf S}_i$ is the spin-$\frac{1}2$ operator at site $i$;  $\langle ij\rangle$ denotes the summation over all the nearest neighbor sties on a honeycomb lattice.  The rotated spin $\tilde {\bf S}_i$ is defined  as follows:  Divide the honeycomb
lattice into four sublattices, keep $\tilde {\bf S}_i={\bf S}_i$
in one sublattice (black) and change  in the other three: $\tilde {\bf
S}_i=(S^x_i,-S^y_i,-S^z_i)$ for the green sublattice, $(-S^x_i,S^y_i,-S^z_i)$  for the grey one and $(-S^x_i,-S^y_i,S^z_i)$ for the red one \cite{khm}. ( See Fig. \ref{fig1}(a).)

\begin{figure}[htb]
\begin{center}
\includegraphics[width=6cm]{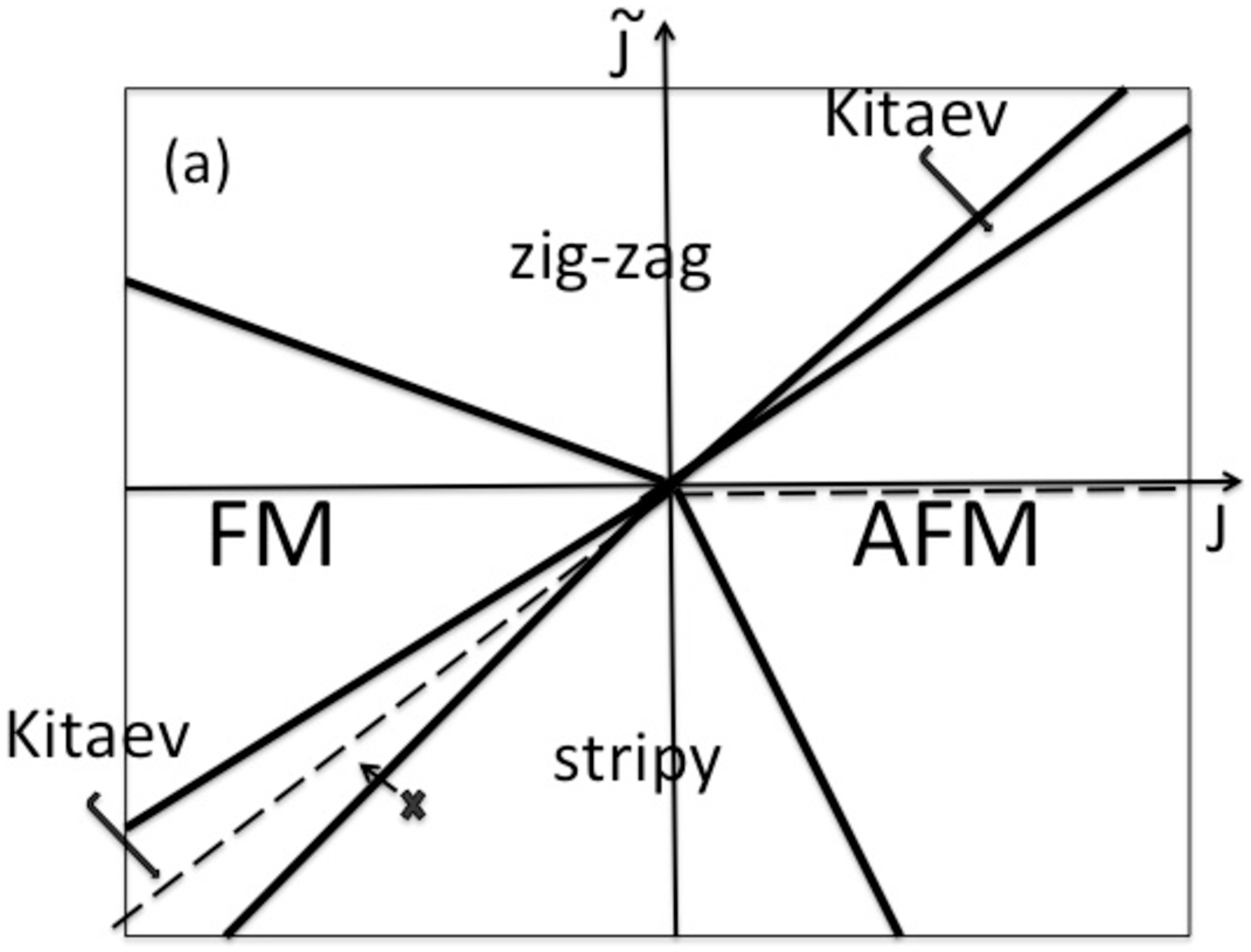}

\vspace{-0.5cm}

\includegraphics[width=6cm]{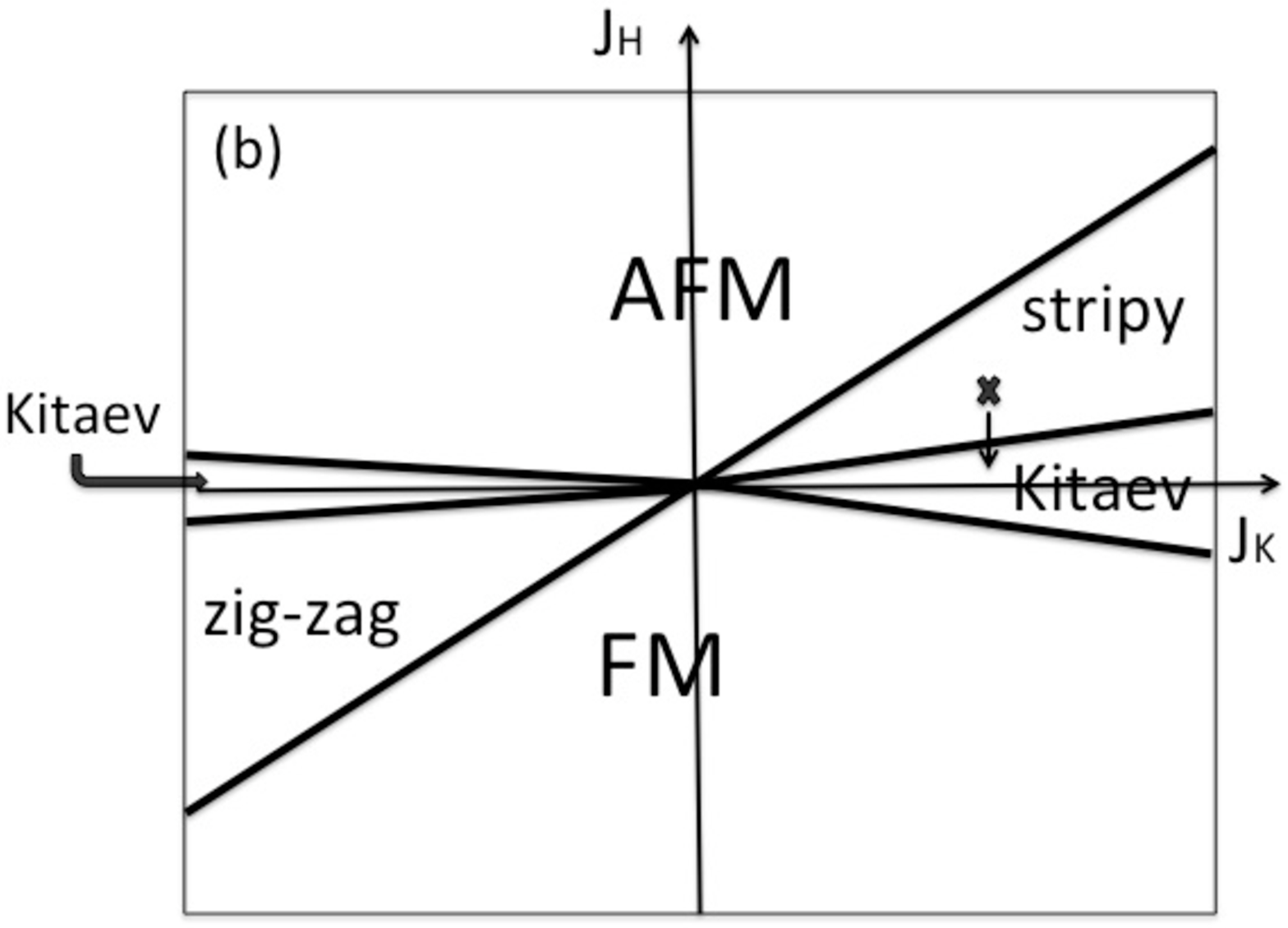}
\vspace{-0.5cm}
\caption{  The phase diagrams : (a) in the $J$-$\tilde J$ plane; (b) in the $J_K$-$J_H$ plane.  The material parameters (the cross symbols) of Li$_2$IrO$_3$ is located in the stripy phase and can be tuned into the Kitaev SL phase by the LDR-SO coupling. This phase diagram was confirmed recently in \cite{12095100}.} \label{fig2}.
\end{center}
\vspace{-0.3cm}
\end{figure}

We then have four magnetic ordered states in the phase diagram: The conventional FM and AFM orders for $\tilde J=0$; and the FM and AFM orders of $\tilde {\bf S}$ for $J=0$, which are the  stripy AFM order  and the zig-zag AFM order of the spin ${\bf S}$ (See Fig.\ref{fig1}), respectively.  The phase structure of the model is governed by the competitions between these ordered magnetic states.  There are phase transitions from the FM (AFM) phase to rotated spin's AFM (FM). The exactly solvable Kitaev model appears at $J=\tilde J$.   The phase diagram is depicted in Fig. \ref{fig2}(a) \cite{yuqin}. 

Defining two new exchange coupling constants 
$J_H=J-\tilde J,~~J_K=-2\tilde J$, the Hamiltonian (\ref{JJ}) reads
\begin{eqnarray}
H=J_H\sum_{\langle ij\rangle}{\bf S}_i\cdot{\bf S}_j-J_K\sum_{a;\langle ij\rangle_a}S^a_iS^a_j
\end{eqnarray}
 where $\langle ij\rangle_a$ runs over all the $a$-links if we distinguish all links of the lattice to be three types of links $a=x,y,z$ (See
 Fig. \ref{fig1}(b).). This is a KH model with arbitrary real number couplings. The phase boundary below dashed lines in Fig.\ref{fig2}(a) now can be read out from the  KH model with $J_H>0$ and $J_K>0$.  The phase boundary in the fourth quarter is given by the line $\beta=\tilde J/J=-2$ according to the result in Ref.\cite{khm}. The phase boundary in the second quarter is then given by exchanging $J$ and $\tilde J$, i.e.,  the line $\beta^{-1}=J/\tilde J=-2$.
The Kitaev SL phase in the third quarter is in between $\beta=3/4$ and 4/3 \cite{khm}. 

 The range of the AFM Kitaev SL phase in the first quarter needs to be determined. A rough estimate is to compare the ground state energies in  both of the Neel
order and the Kitaev SL.  The
nearest neighbor spin-spin correlation in the Neel order is
$\langle {\bf S}_i\cdot{\bf S}_j\rangle\approx -0.37$ \cite{khm}.
Using this correlation, one can estimate the ground state energy
per site for the Neel order
$e_N/J\approx(1-\beta/3)\langle{\bf S}_i\cdot{\bf S}_j\rangle$.  The upper boundary of  the ground state
energy per site for  the Kitaev SL is given by
$e_K/J\lesssim (1+\beta)\langle S_i^aS_j^a\rangle$ for
$\langle S_i^aS^a_j\rangle=-0.13$ \cite{ssc}. The critical
point can be estimated by comparing the energies 
per site of the two states: $e_N=e_K$,  which requires $\beta=0.95$. The critical values are,
therefore, $\beta_c\lesssim 0.95$ and   $\beta^{-1}_c
\gtrsim1.05$. 

In the positive $J_K$ and $J_H$ KH model, the critical points of the phase transition were determined by calculating the ground state expectation values of a group of physical observables, e.g., the square of total spin, the nearest neighbor spin correlations, and the second derivative of the ground state energy for the parameter $\beta$ in exact diagonalization method \cite{khm}.
We also calculated these quantities by exact diagonalization calculations. Unfortunately,  up to 24 lattice sites,
although we can see that all these quantities are smooth for $\beta< 0.9$ and dramatically change in  $\beta\in$[0.9,1.0],  we cannot determine a sharp critical value of $\beta$ according to the data for these quantities. 

Instead of examining  the ground state, we focus on the first excited state.  In mean field approximation, the low-lying excitation in the Neel order is a gapless bosonic spin wave.  The dispersion of the spin wave in terms of the Hostein-Primakoff transformation is given by  $J\sqrt{[(3-\beta+2\beta^2)q_x^2+3(1-\beta)q_y^2]/8}$ for $|{\bf q}|\to 0$. However, these spin wave excitations are in fact gapped  due to the lacking of the SU(2) symmetry away from either $J=0$ or $\tilde J=0$ \cite{khm}.  On the other hand, the fermonic gapless excitation in the Kitaev SL phase can not be gapped due to protection of the time reversal symmetry and is
a linear combination of the A and B sublattice Majorana fermion
modes. The dispersion of this Majorana fermion excitation obtained
by a mean field theory is of the form $ \sqrt{3}(2- \beta)J|{\bf q}|$ near the Dirac points ${\bf K}_\pm=(\pm \frac{4\pi}3,0)$.

\begin{figure}[htb]
\begin{center}
\includegraphics[width=6cm]{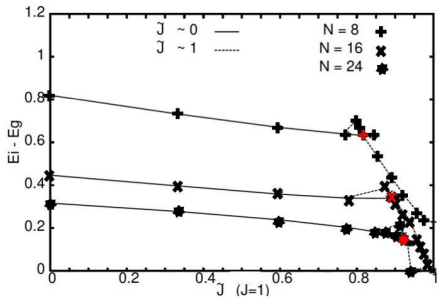}
\caption{ (Color online) The energy levels of the first excited states (subtracting the ground state energy) for lattice size    $N=8,16$ and 24.  The level crossing points (red symbles)  determine the critical point. The solid lines labelled as $\tilde J\sim 0$ connect the calculation data starting from $\tilde J=0$ (the Neel order). The dashed lines ($\tilde J\sim1$) connect the data  starting from $\tilde J=J$ (the Kitaev model). }\label{fig3}
\end{center}
\vspace{-0.2cm}
\end{figure}

For a finite system, the difference between
 these first excited states can be revealed by considering
 the variation of their excitation energies as $\beta$.
 The critical points can be determined by looking for the level crossing of the first excitation state energy levels of two different phases in a finite size (See Fig. \ref{fig3}). Up to 24-sites, the data  from exact diagonalization gives $\beta_c\approx 0.93$ for the phase transition from the Neel order to the Kitaev SL phase which is close to our estimate $\beta_c\lesssim 0.95$.  This is also consistent with recent result in \cite{12095100}.

In Fig. \ref{fig2}(b), we redraw the phase diagram in $J_K$-$J_H$ plane. The FM Kitaev SL phase is  in $|\gamma|=|J_H/J_K|<0.1$ and the AFM Kitaev SL phase is in $|\gamma|<0.038$; The stripy AFM in $0.01<\gamma<0.75$; and the zig-zag in  $0.038<\gamma<0.75$.

\noindent{\it Possible microscopic origins of the $J$-$\tilde J$ model. }  We discuss possible microscopic origins of the $J$-$\tilde J$ exchange couplings. We restrict our study to a single band Hubbard model.  For electron $c_{i\sigma}$ on the lattice, spin-dependent hopping $ -t_s c^\dag_{i\sigma}\sigma^a_{\sigma\sigma'}  c_{j\sigma'}$ for a given $a$- link may cause a rotated spin AFM exchange in large $U$ limit. (Here $\sigma^a$ is Pauli matrices. For a brief discussion, see below.)   Replacing the fermion by two-component boson, one can arrive at  the
FM $J$-$\tilde J$ model. However, the spin-dependent hopping is not easy to be manipulated in condensed matter systems. (It is possible to be realized the spin-dependent hopping in cold atom systems and we will  study this matter elsewhere with respect to cold atom physics. )

\begin{figure}[htb]
\includegraphics[width=5.5cm]{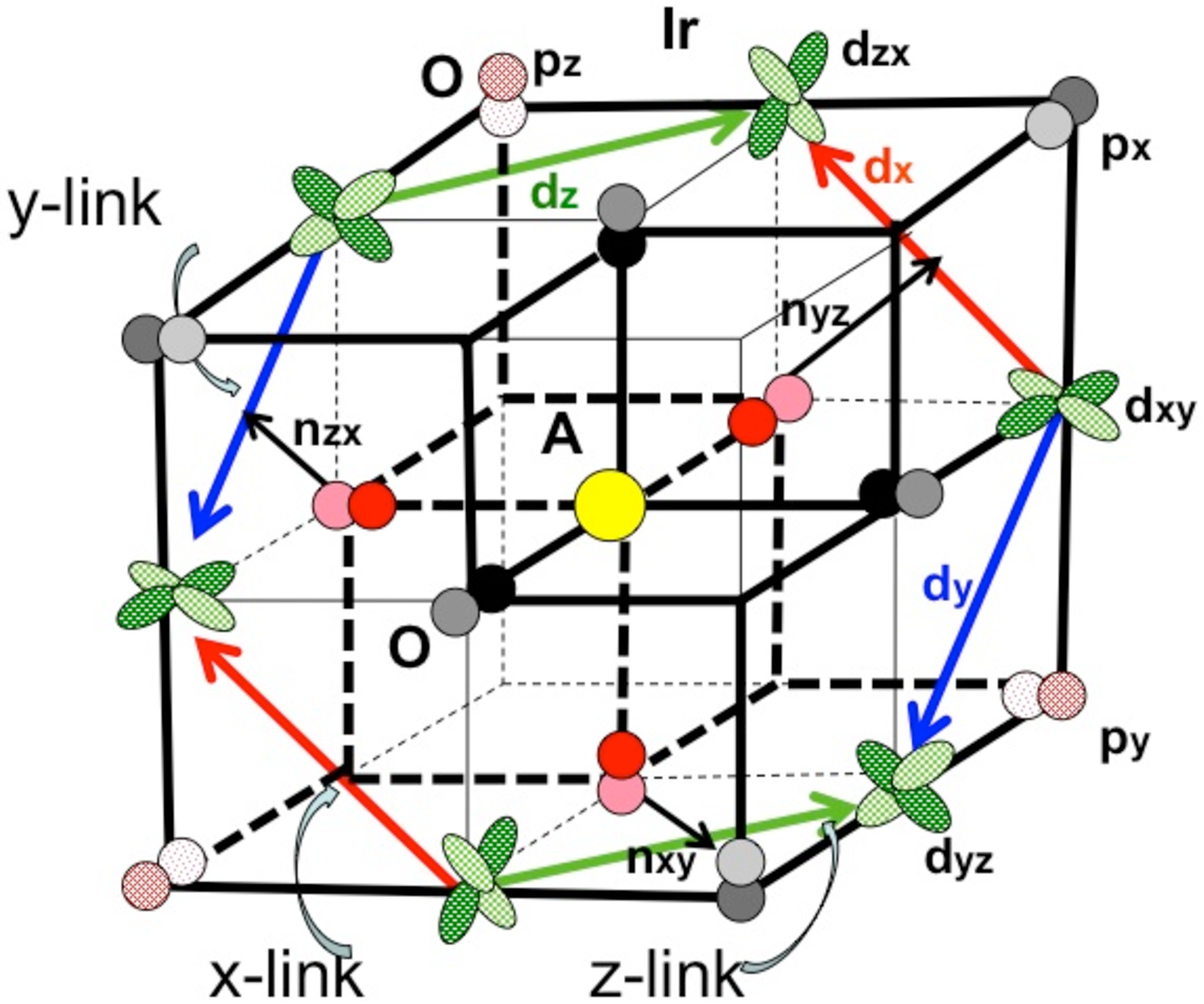}
\vspace{-3mm}
\includegraphics[width=5.5cm]{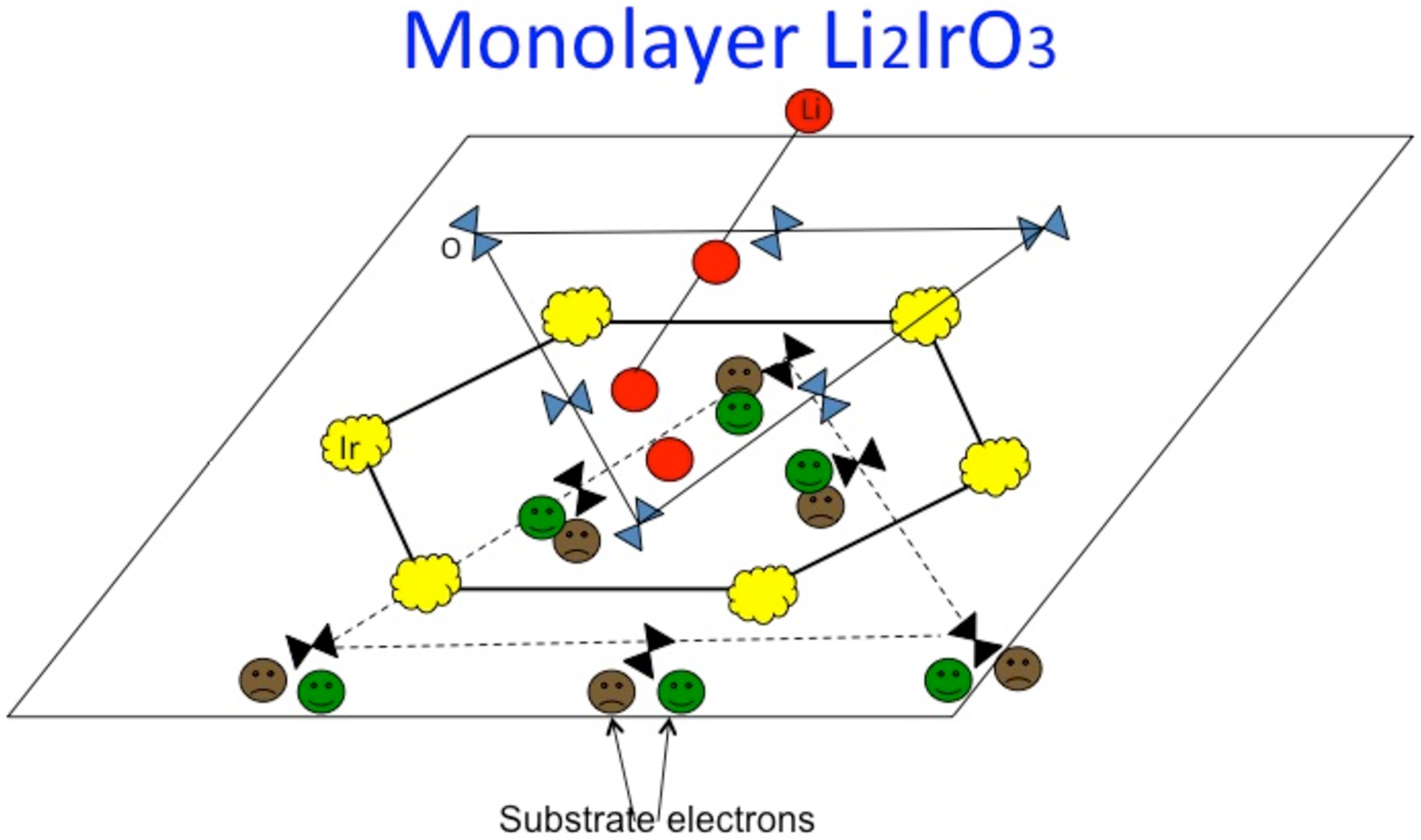}
\caption{(Color online) Top: AIr$_2$O$_6$ layer in A$_2$IrO$_3$ ($d_{ab}$ states for Ir$^{4+}$ ). The superexchange orbital configuration of Ir$^{4+}$ ($d_{ab}$) and  O$^{2-}$ ($p^a$) are depicted at the position of these ions. The white-grey triangular lattice of oxygens is on the top of the honeycomb lattice of indiums ; the red-pink one is  beneath the honeycomb. The vectors ${\bf n}_{xy}$=[110]; ${\bf d}_z=[-110]$  etc. Bottom: Monolayer of A$_3$ (AIr$_2$O$_6$) on substrate.  }\label{fig4}
\end{figure}

 Relevant to the iridates, we consider LDR-SO couplings.   Fig. \ref{fig4}(a)  shows an AIr$_2$O$_6$ layer of  A$_2$IrO$_3$.  The effective spin-$\frac{1}2$ (J$_{\rm eff}=\frac{1}2$  ) $5d$-electron gas on the honeycomb lattice of Ir$^{4+}$ behaves like a Mott insulator\cite{jeff}. We consider a monolayer of the compound  (Fig. \ref{fig4}(b)) which is on the top of a substrate whose lattice structure matches the triangular lattice of the oxygens (black $p$-orbitals) beneath the honeycomb lattice.  Due to the influence of the substrate, the interaction between the effective spin-$\frac{1}2$ electrons and the $p$-orbital states of the oxygens in the black triangular lattice is different from that with the $p$-electrons of the oxygens in the blue triangular lattice on the top of the honeycomb lattice. This asymmetry induces a field whose local direction is ${\bf n}_{ab}$  and  then a LDR-SO interaction, e.g.,  $i\lambda_R{\bf d}_z\times\vec
\sigma\cdot {\bf n}_{xy}$  and its xyz cyclic permutations, where the dipole moment ${\bf d}_z$ is parallel to the electron hopping direction, and $\lambda_R$ is the  Rashba SO
coupling strength. We identify the links labelled by ${\bf d}_{ij}={\bf
d}_a$ with the $a$-links. In second quantization,  this 
reads $i\lambda_R\nu_{ij}c^\dag_{i\sigma}
\sigma^a_{\sigma\sigma'}c_{j\sigma'}$ for link $a$ ($i,j$
label the lattice sites connected this link) where
$\nu_{ij}=-\nu_{ji}=1$ since ${\bf d}_{ij}=-{\bf d}_{ji}={\bf
d}_a$.

\noindent{\it The LDR-SO Hubbard model and the large U limit.} The Hamiltonian of the
LDR-SO Hubbard model on honeycomb lattice reads
\begin{eqnarray}
H_{\rm LDR-SO}&=&H_T+H_R+H_U
=-t\sum_{\langle ij\rangle}c^\dag_{i\sigma}c_{j\sigma}\\
&+&i\lambda_R\sum_{a;\langle ij\rangle_a}\nu_{ij}c^\dag_{i\sigma}
\sigma^a_{\sigma\sigma'}c_{j\sigma'}+U\sum_{i}n_{i\uparrow}n_{i\downarrow}\nonumber
\end{eqnarray}
where the $J_{\rm eff}$=1/2 'spin' is labelled by $\sigma=\uparrow,\downarrow$; $H_T$ is the direct electron hopping between iridium sites; $H_R$ is the external LDR-SO coupling and $H_U$ is on-site Hubbard repulsion.

Take $H_T$ and $H_R$ as the perturbations in the large $U$ limit. To second order for a given link $a$ with ${\bf d}_{ij}=\nu_{ij}{\bf d}_a$, the effective Hamiltonian for the half filling case is given by,
\begin{eqnarray}
&-&\frac{2}U[(H_T)_{ij}(H_T)_{ji}+(H_R)_{a,ij}(H_R)_{a,ji}\nonumber\\
&+&(H_R)_{a,ij}(H_T)_{ji}
+(H_T)_{ij}(H_R)_{a,ji}=J_t{\bf S}_i\cdot{\bf S}_j\\&+&J_R[S^a_iS^a_j- S_i^bS^b_j-S^c_iS^c_j]
+2\nu_{ij}J_{tR}(S^b_iS^c_j-S^c_iS^b_j)\nonumber
\end{eqnarray}
where  $J_t=\frac{4t^2}U$, $J_R=\frac{4\lambda_R^2}U$ and  $J_{tR}=\frac{4t\lambda_R}U$
;  $a\ne b\ne c$  (e.g., $a=x, b=y$ and $c=z$ etc.) and ${\bf S}=\frac{1}2c^\dag_\sigma{\vec\sigma}_{\sigma\sigma'}c_{\sigma'}$.  Compactly, the LDR-SO Hubbard model reduces to
\begin{eqnarray}
H_{t\lambda}&=&J_t\sum_{\langle ij\rangle}{\bf S}_i\cdot{\bf S}_j+J_R\sum_{\langle ij\rangle}\tilde{\bf S}_i\cdot\tilde{\bf S}_j\label{tR}
\nonumber\\&+&2J_{tR}\sum_{a,\langle ij\rangle_a}({\bf n}_{bc}\times {\bf d}_{ij})\cdot{\bf S}_i\times {\bf S}_j,
\end{eqnarray}
where the last term is a link-dependent Dzyaloshiskii-Moriya coupling. (This term comes from the cross term  between $H_T$ and $H_R$.  Notice that the cross term between $H_T$ and the spin-dependent hopping $-t_s\sum_{a;\langle ij\rangle_a}c^\dag_{i\sigma}\sigma^a_{\sigma\sigma'}c_{j\sigma'}$ vanishes. Therefore,  the $J$-$\tilde J$ model can precisely describe the strong coupling physics of the Hubbard model with spin-dependent hoppings.)

\noindent{\it Tuning into Kitaev SL phase in the  iridates.}   The zig-zag AFM order was found in experiments for the iridates \cite{liu,singh,choi,cao}.  This zig-zag order phase in the phase diagram appears in the third quarter of Fig. \ref{fig2}(b), which might originate from the inter-orbital  $t_{2g}$-$e_g$ hopping \cite{12095100}. Another origin of the zig-zag order is from the contribution of  $J_2$ and $J_3$ Heisenberg couplings to the positive $J_K$ and $J_H$ KH model by considering the intrinsic SO coupling \cite{KY}.   As discussed before, for a monolayer of the iridates, the LDR-SO coupling needs to be added and then the two-dimensional strong coupling Hamiltonian reads  
\begin{eqnarray}
H_{\rm Ir}=J_H\sum_{\langle ij\rangle}{\bf S}_i\cdot{\bf S}_j+H_{KDM}+[J_2]+[J_3],\label{J_K1}\\
H_{\rm KDM}=- J_K\sum_{a;\langle ij\rangle_a}[S^a_iS^a_j+\nu_{ij}\gamma_{tR}(S^b_iS^c_j-S^c_iS^b_j)]\label{eff}
\end{eqnarray}
where $J_H=J_t-J_R$, $J_K=J_{in}-2J_R$ and $\gamma_{tR}=\frac{2J_{tR}}{J_K}$.  $[J_2]$ and $[J_3]$ denote the second and third nearest neighbor Heisenberg exchange terms  \cite{KY} as they are not negligible in A$_2$IrO$_3$ \cite{liu,singh,choi,cao}. 
$H_{KDM}$ is no longer exactly solvable. With Majorana fermion's representation \cite{kitaev}, i.e., $(b^x)^2=(b^y)^2=(b^z)^2=c^2=1$ as well as anti-commutation between different fermions, we have
\begin{eqnarray}
H_{\rm KDM}=iJ_K\sum_{a;\langle ij\rangle_a}(ib^a_ib^a_j+\nu_{ij}\gamma_{tR}(ib^b_ib^c_j-ib^c_ib^b_j))c_ic_j.
\end{eqnarray}
A mean field theory with parameters $\langle(ib^a_ib^a_j+i\nu_{ij}\gamma_{tR}(b^b_ib^c_j-b^c_ib^b_j))\rangle$ and $\langle c_ic_j\rangle$ decouples $c$-fermions and $b$-fermions. While the  $c$-fermions give 
Kitaev model, the $b$-fermions are gapped if $\gamma_{tR}<1$. Therefore,  $H_{KDM}$  with $\gamma_{tR}<1$ can adiabatically drive to  Kitaev model ($\gamma_{tR}=0$).  In other words, when $H_{KDM}$ dominates, the system is in the Kitaev SL phase. If $\gamma_{tR}\geq 1$, $b$-fermions become gapless, e.g., for $\gamma_{tR}=1$, $b$-fermions are gapless at Dirac points $(\pm\frac{4\pi}3,0),(0,\pm\frac{2\pi}{\sqrt3}), (\pm\pi,\pm\frac{\pi}{\sqrt3})$.  This will not be relevant  as we will see.

We parameterize the Hamiltonian (\ref{J_K1}) by $J_H=(1-\alpha)A$ and $J_K=2\alpha A$ where $A=(J_K+2J_H)/2$ \cite{khm}.   Define $\alpha_{ir}=J_{in}/(J_{in}+2J_t)$  which is a material parameter of the iridates . Comparing with the material parameter $\alpha_{ir}$,  the model controlling parameter  $\alpha$  is enhanced, i.e.,   
 \begin{eqnarray}
\alpha=\alpha_{ir}\frac{1-\beta_{Rin}}{1-2\alpha_{ir}\beta_{Rin}}
\end{eqnarray}
where $\beta_{Rin}=2J_R/J_{in}<1$.  There is a divergence at  $2\alpha_{ir} \beta_{Rin}=1$  if $\alpha_{ir}>0.5$.  Thus,  $\alpha$ can be tuned to exceed any finite critical value.  

For $J_t$ and $J_{in}>0$, the typical values of $J_t\sim 5$meV,  $U\sim 3$eV, and  $\alpha_{ir}\gtrsim0.65$ in the bulk Li$_2$IrO$_3$ \cite{singh}.  Assuming that the bulk parameters is still valid for the monolayer iridate,  $\alpha$ may be tuned into the Kiatev phase (The cross symbols in Fig. \ref{fig2}). In Table I (first two lines), we list the critical values $\alpha_c$, the corresponding values of $J_H/J_K$, $\beta_{Rin}$, $\lambda_R$, $J_R$ and $\gamma_{tR}$ to $\alpha_c$ for Li$_2$IrO$_3$ with two sets of different $J_2$ and $J_3$. ( Since the $b$-fermions are gapped, the correction to $\alpha_c$ from them is negligible. )
\begin{center}
 \vline\vline\vline \begin{tabular}{ c| c|c|c|c|c|c|c }
  \hline\hline
    $\frac{J_2}{J_H}$&$\frac{J_3}{J_H}$&$\alpha_c$&$(\frac{J_H}{J_K})_c$&$(\beta_{Rin})_c$&$(\lambda_R)_c$&$(J_R)_c $&$\gamma_{tR}$\\ \hline
    0&0& 0.8 & 0.125&0.38&51.45&3.53&0.57\\ \hline
   0.6&0.6 &  0.83&0.102&0.42&54.08&3.90&0.62\\\hline
   0&0&0.93&0.038&0.50&-&0.93$J_t$&-\\
    \hline\hline
  \end{tabular}\vline\vline\vline
  \end{center} 
  
   \vspace{1mm}
  \noindent{\small Table I: The critical value $\alpha_c$ for $J_2=J_3=0.6$ is from Ref.\cite{singh}. 
  The unit of $(\lambda_R)_c$ and $(J_R)_c$ is meV.  }
\vspace{1mm}

  The critical LDR-SO coupling strength is of the order $(\lambda_R)_c\sim 50$meV which is about  a factor $1/10$ weaker than the intrinsic SO coupling ($\lesssim 500$meV).  $\gamma_{tR}<1$ insures the SL phase is the Kitaev SL. 
  
   For $J_t$ and $J_{in}<0$, i.e., the $t_{2g}$-$e_g$ inter-orbital coupling is taken into account \cite{12095100}, it is also possible to tune into the Kitaev SL phase as shown in last line of Table I if one takes Eq. (\ref{eff}) as the effective model and $\alpha_{in}=0.65$. However, the microscopic origin of $J_R$ is not studied in this case.
  
For Na$_2$IrO$_3$, due to a small $\alpha_{in}\lesssim0.25$ \cite{singh}, $\alpha$ can only be tuned to a smaller value $0<\alpha\lesssim0.25$ by a weak  $J_R$, i.e.,  the system moves to deeper magnetic ordered state by the LDR-SO coupling.

In conclusion,  we have studied the phase diagram of  the $J$-$\tilde J$ model. We showed that the Kitaev SL phase can be a result of  competition between two Landau-type ordered states. We studied the microscopic origin of this $J$-$\tilde J$ model within a single band Hubbard model.  We showed that this model can be applied to the iridates by introducing the LDR-SO coupling.  We suggest to put a monolayer  Li$_2$IrO$_3$ on top of a substrate and predict that the Kitaev phase is 
 reachable by tuning the LRD-SO coupling.  

This work was supported by  the 973 program of MOST
of China (2012CB821402, 2009CB929101), NNSF of China and DOE of USA (DE-FG03-02ER45958).

\end{document}